\documentclass[10pt]{amsart}

\usepackage{color}

\def\t{\mbox{\boldmath$t$}}
\def\fb{\mbox{\boldmath$f$}}
\def\x{\mbox{\boldmath$x$}}
\def\bbeta{\mbox{\boldmath$\beta$}}

\begin{document}

\title[Critical points, Lauricella functions and
Whitham-type equations]{Critical points, Lauricella functions and
Whitham-type equations}

\author{Y. Kodama$^1$, B. Konopelchenko$^2$ and W.K. Schief$^3$}
\address{$^1$Department of Mathematics, Ohio State University, Columbus,
OH 43210}
\address{$^2$Department of Mathematics and Physics ``Ennio de Giorgi'', University of Salento and sezione INFN, Lecce, 73100, Italy}
\address{$^3$School of Mathematics and Statistics and Australian Research Council Centre of Excellence for Mathematics and Statistics of Complex Systems, The University of New South Wales, Sydney, NSW 2052, Australia}
 
\begin{abstract}
A large class of semi-Hamiltonian systems of hydrodynamic type is
interpreted as the equations governing families of critical points of
functions obeying the classical linear Darboux equations for conjugate nets. The
distinguished role of the Euler-Poisson-Darboux equations and associated
Lauricella-type functions is emphasised. In particular, it is shown that the
classical $g$-phase Whitham equations for the KdV and NLS equations are
obtained via a $g$-fold iterated Darboux-type transformation generated by appropriate
Lauricella functions.
\end{abstract}

\maketitle


\section{Introduction}
The Whitham equations and other semi-Hamiltonian systems of hydrodynamic type represent an important subclass of diagonal systems of first-order
quasi-linear partial differential equations
\begin{equation}\label{eq:1}
\frac{\partial \beta_i}{\partial t_{2}}=\lambda _{i}(\beta _1,\ldots,\beta_N)\frac{\partial \beta _i}{\partial t_{1}},\qquad i=1,\ldots,N
\end{equation}
which exhibits rich mathematical structure and has a variety of applications in physics (see,
e.g., \cite{Wh, RY}).  The theory of Hamiltonian systems of hydrodynamic type has been developed by Dubrovin and Novikov \cite{DN}. Intimate connections with the theory of conjugate and orthogonal nets of classical differential geometry have been established in \cite{Per70,Gru84,T1,T2}. The construction of Whitham-type equations, that is, the determination of the associated characteristic speeds, usually relies heavily on the sophisticated theory of Riemann surfaces \cite{FFM, FL, Kr1, Kr2}. By contrast, the approach presented in this paper is based on the interpretation of the system
\eqref{eq:1} as a system of partial differential equations describing deformations of critical points and the classical theory of Darboux-type transformations for conjugate nets.

For any function $\Theta(x_1,\ldots,x_N;t_1,\ldots,t_n)$, the critical points
$\bbeta=(\beta_1,\ldots,\beta_N)$ of $\Theta$ are defined implicitly by
\begin{equation}\label{E1}
\left.\frac{\partial \Theta}{\partial x_i}\right|_{\x=\bbeta}=0,\qquad i=1,\ldots,N.
\end{equation}
In the current context, the function $\Theta(\x;\t)$ is assumed to be a solution of the system of linear equations
\begin{equation}\label{eq:2}
\frac{\partial^2\Theta}{\partial x_{i}\partial x_{k}}=A_{ik}\frac{\partial\Theta}{\partial x_{i}}+A_{ki}\frac{\partial \Theta}{\partial x_{k}},\qquad i\neq k
\end{equation}%
for some functions $A_{ik}(\x)$ and any $\t$.   The set of equations \eqref{eq:2} is well-known in the classical theory of conjugate nets \cite{Darboux}. Particular choices 
of solutions $\Theta $ of the system \eqref{eq:2} give rise to different
semi-Hamiltonian systems of the type \eqref{eq:1}. The latter turn out to be invariant under the classical Combescure transformation for conjugate nets acting on $\Theta$. A privileged subclass of systems \eqref{eq:2} is given by the choice
\begin{equation}
A_{ik}=\frac{\varepsilon_k}{x_i-x_k},
\end{equation}
leading to the Euler-Poisson-Darboux (EPD) equations 
\begin{equation}\label{eq:3}
\frac{\partial f}{\partial x_{i}\partial x_{k}}=\frac{1}{x_{i}-x_{k}}\left(\varepsilon _{k}\frac{\partial f}{\partial x_{i}}-\varepsilon
_{i}\frac{\partial f}{\partial x_{k}}\right),\qquad i\neq k
\end{equation}%
with constants $\varepsilon _{1},\ldots,\varepsilon _{N}$.
The latter system of equations is seen to play a distinguished role in our approach. The classical Fundamental, Levy and other transformations (see, e.g., \cite{Eisenh}) acting on the system \eqref{eq:3} allow us to construct large classes of systems \eqref{eq:2} and their solutions. In particular, it is shown that the characteristic speeds $\lambda _{i}$ associated with the standard $g$-phase Whitham equations for the Korteweg-de Vries (KdV) and nonlinear Schr\"odinger (NLS) equations may be obtained via iterated Darboux-type
transformations generated by Lauricella-type solutions (specific complete
hyperelliptic integrals) \cite{Lau} of an extended EPD system \eqref{eq:3} with all $
\varepsilon _{i}=\frac{1}{2}$. It is noted that, in the construction presented here, the sophisticated theory of Riemann surfaces need not be employed.

The general idea that (classes of) nonlinear partial differential equations may be regarded as the stationary (critical) points of functionals (functions) is classical and has been elaborated upon in a variety of contexts. In this paper, it is demonstrated that the notion of critical points is natural and may be used very effectively in the specific context of hydrodynamic-type systems \eqref{eq:1}. It is also emphasised that even though the significance of the EPD equations in the theory of Whitham equations has been established
in \cite{KS1, KS2, GKE, Tian}, their role is quite different in the approach presented here.

\section{Semi-Hamiltonian hydrodynamic type systems}

In the seminal paper \cite{T1}, Tsarev proved that, at least locally, all
solutions $\bbeta=\bbeta(t_1,t_2)$ of a semi-Hamiltonian system \eqref{eq:1} are implicitly given by the
system
\begin{equation}\label{eq:4}
t_{1}+\lambda _{i}(\bbeta )t_{2}-\omega _{i}(\bbeta )=0,\qquad
i=1,\ldots,N,
\end{equation}%
where the set of functions $\left\{ \omega_{i}\right\} $ is the solution of the
linear equations
\begin{equation}\label{eq:5}
\frac{\partial \omega _{i}}{\partial \beta _{k}} = B_{ik}(\omega _{k}-\omega_{i}),\qquad i\neq k
\end{equation}%
with the coefficients $B_{ik}$ being defined by
\begin{equation}\label{E2}
\frac{\partial \lambda _{i}}{\partial \beta _{k}} = B_{ik}(\lambda_k - \lambda_i).
\end{equation}
The condition for the existence of the functions $\omega_i$ (in the sense of a natural Cauchy problem) is equivalent to the assumption that the system (\ref{eq:1}) be semi-Hamiltonian and may be expressed as
\begin{equation}\label{E3}
\frac{\partial B_{ik}}{\partial\beta_l} = \frac{\partial B_{il}}{\partial\beta_k},\qquad i\neq k\neq l\neq i.
\end{equation}
The latter together with the compatibility conditions $\frac{\partial}{\partial\beta_l} (\frac{\partial \lambda_i}{\partial\beta_k})=\frac{\partial}{\partial\beta_k} (\frac{\partial \lambda_i}{\partial\beta_l})$
give rise to the first-order system
\begin{equation}\label{E4}
\frac{\partial B_{ik}}{\partial \beta_l} = B_{ik}B_{kl} - B_{ik}B_{il} + B_{il}B_{lk}.
\end{equation}

Semi-Hamiltonian systems admit an infinite set of
commuting symmetries given by
\begin{equation}\label{eq:6}
\frac{\partial \beta _i}{\partial t_{\alpha }}=\lambda _{i}^{\alpha }(\bbeta )\frac{\partial \beta _i}{\partial t_{1}},\qquad\alpha =2,3,\ldots,
\end{equation}%
where each set $\{\lambda_i^\alpha\}$ necessarily constitutes another solution of the linear system (\ref{E2}) due to the compatibility condition associated with (\ref{eq:6}). Here, we have made the identification $\lambda_i^2 = \lambda_i$.
The symmetries, in turn, give rise to an infinite family of conservation laws
\begin{equation}\label{E4b}
\frac{\partial P}{\partial t_{\alpha }}=\frac{\partial Q^\alpha}{\partial t_1},
\end{equation}
where the density $P$ and the fluxes $Q^\alpha$ are related by
\begin{equation}\label{eq:8}
\frac{\partial Q^{\alpha}}{\partial \beta _{i}}=\lambda _{i}^{\alpha }(\bbeta )\frac{\partial P}{\partial\beta _{i}}.
\end{equation}%
The existence of the fluxes is guaranteed if the density obeys the system of linear partial differential equations \cite{T1}
\begin{equation}\label{eq:9}
\frac{\partial P}{\partial\beta _{i}\partial \beta _{k}}=B_{ik}\frac{\partial P}{\partial\beta _{i}}+B_{ki}\frac{\partial P}{\partial \beta _{k}},\qquad i\neq k.
\end{equation}%
These are indeed compatible modulo the first-order system (\ref{E4}). In fact, the above linear system represents the basic equations of the theory of $N$-dimensional conjugate nets developed more than a century ago \cite{Dar}. In this context, the connection between any quantity $Q^\alpha$ and $P$ is known as a Combescure transformation (see, e.g., \cite{RogersSchief2002}). Moreover, the parametrisation 
\begin{equation}
B_{ik}=\frac{\partial}{\partial \beta_k}\ln B_{i}
\end{equation}
implied by the ``conservation laws'' (\ref{E3}) leads to the integrable nonlinear Darboux system (see, e.g., \cite{Kon93}) encapsulated in (\ref{E4}). The relevance of conjugate nets, not necessarily of orthogonal (such as Egoroff) type, in the theory of semi-Hamiltonian systems has been noted in \cite{T2}.

A natural extension of Tsarev's theorem for commuting systems of hydrodynamic type is readily available. Thus, all (local) solutions $\bbeta = \bbeta(\t)$ of $n-1$ commuting flows (\ref{eq:6}) are given implicitly by the system
\begin{equation}\label{eq:10}
\sum_{\alpha=0}^n t_{\alpha }\lambda _{i}^{\alpha }(\bbeta )=0,\qquad
i=1,\ldots,N,
\end{equation}
where 
\begin{equation}
 t_0 = 1,\qquad\lambda_i^1=1
\end{equation}
and $\{\lambda_i^0\}$ is the ``general solution'' of the linear system (\ref{E2}). This may be proven in the same manner as Tsarev's theorem for $n=2$ \cite{T1}. 

The system of equations (\ref{eq:10}) possesses a distinguished property. Indeed, by virtue of the relations \eqref{eq:8}, it may be reformulated as
\begin{equation}\label{eq:11}
\sum_{\alpha=0}^nt_{\alpha }\frac{\partial Q^\alpha}{\partial \beta _{i}}=0,
\end{equation}
where $Q^0$ represents the ``flux'' associated with the set of functions $\{\lambda_i^0\}$ and $Q^1 = P$. Accordingly, if we introduce the parameter-dependent function
\begin{equation}\label{eq:12}
\Theta(\x;\t) =\sum_{\alpha=0}^nt_{\alpha }Q^{\alpha }(\x)
\end{equation}
then the criticality conditions
\begin{equation}
\left.\frac{\partial \Theta}{\partial x_i}\right |_{\x=\bbeta}=0,\qquad i=1,\ldots,N
\end{equation}
coincide with (\ref{eq:11}). Hence, the generalised Tsarev equations \eqref{eq:10} are descriptive of the critical points of the family of functions (\ref{eq:12}). We therefore conclude that  the theory of semi-Hamiltonian hydrodynamic-type systems of the form \eqref{eq:1} is essentially encoded in the theory of conjugate nets via the linear system \eqref{eq:9}, Combescure transformations \eqref{eq:8} and critical points of the family of functions \eqref{eq:12}. It is observed that Hamiltonian systems of the form \eqref{eq:1} may be viewed as the equations for critical (stationary) points of suitable functions. This fact may be traced back to the original derivation by Whitham \cite{Wh} of the equations for slow modulations (stationary phase) and has subsequently been discussed in different forms in \cite{Dubr1, Lo, KMM}.

In order to reveal the key structure behind the connection between semi-Hamil\-tonian hydrodynamic-type systems and critical points, we observe that elimination of the density $P$ in (\ref{eq:8}) via cross-differentiation shows that the fluxes $Q^\alpha$ are likewise solutions of conjugate net equations with coefficients depending on $\alpha$. However, on use of (\ref{eq:9}), differentiation of (\ref{eq:8}) produces
\begin{equation}\label{E4a}
 \frac{\partial Q^\alpha}{\partial\beta _{i}\partial \beta _{k}}=A_{ik}\frac{\partial Q^\alpha}{\partial\beta _{i}}+A_{ki}\frac{\partial Q^\alpha}{\partial \beta _{k}},\qquad i\neq k
\end{equation}
with coefficients
\begin{equation}
A_{ik} = B_{ki}\left.\frac{\partial P}{\partial \beta_k}\right/\frac{\partial P}{\partial \beta_i}
\end{equation}
which are independent of $\alpha$. Hence, $\Theta$ as defined by (\ref{eq:12}) may be regarded as a linear superposition of solutions of the {\em same} conjugate net equations (\ref{E4a}). It is noted that
\begin{equation}
\frac{\partial A_{ik}}{\partial\beta_l} = \frac{\partial A_{il}}{\partial\beta_k},\qquad i\neq k\neq l\neq i
\end{equation}
and, hence, the compatibility conditions for the linear system (\ref{E4a}) imply that the coefficients $A_{ik}$ constitute another solution of the first-order system (\ref{E4}).

\section{General scheme}

In order to formalise the observations made in the previous section, it is natural to consider functions $\Theta(\x)$ for which the associated Hessian matrix of second-order derivatives $\frac{\partial^2\Theta}{\partial x_i\partial x_k}$ is diagonal at any critical point $\x=\bbeta$. In the linear case, this condition may be met by demanding that $\Theta$ be a solution of a system of conjugate net equations
\begin{equation}\label{eq:13}
\frac{\partial^2\Theta}{\partial x_{i}\partial x_{k}}=A_{ik}\frac{\partial\Theta}{\partial x_{i}}+A_{ki}\frac{\partial\Theta}{\partial x_{k}},\qquad
i\neq k \in\{1,\ldots, N\}.
\end{equation}%
Indeed, if the criticality conditions are satisfied then the mixed derivatives of $\Theta$ vanish, that is,
\begin{equation}
\left.\frac{\partial \Theta}{\partial x_i}\right |_{\x=\bbeta}=0\qquad\Rightarrow\qquad \left.\frac{\partial^2 \Theta}{\partial x_i\partial x_k}\right |_{\x=\bbeta}=0,\qquad i\neq k.
\end{equation}
In this connection, it is convenient to introduce the notation
\begin{equation}
f_{\beta_i} = \left.\frac{\partial f}{\partial x_i}\right|_{\x=\bbeta}
\end{equation}
for any function $f = f(\x)$ so that the criticality conditions may be formulated as $\Theta_{\beta_i}=0$. As mentioned in the preceding, the system (\ref{eq:13}) is compatible if the coefficients $A_{ik}$ obey the first-order equations
\begin{equation}\label{darboux}
\frac{\partial A_{ik}}{\partial x_l} = A_{ik}A_{kl} - A_{ik}A_{il} + A_{il}A_{lk}.
\end{equation}
The latter imply that
\begin{equation}
A_{ik} = \frac{\partial}{\partial x_k}\ln A_i
\end{equation}
for some potentials $A_i$ so that the nonlinear Darboux system
\begin{equation}\label{eq:14}
\frac{\partial^2 A_i}{\partial x_{k}\partial x_{l}}=\frac{1}{A_k}\frac{\partial A_k}{\partial x_l}\frac{\partial A_i}{\partial x_{k}}+\frac{1}{A_l}\frac{\partial A_l}{\partial x_k}\frac{\partial A_i}{\partial x_{l}},\qquad i\neq k\neq l\neq i
\end{equation}
results.

In order to study critical points which depend on some parameters $\t = (t_1,\dots,t_n)$, we consider the linear combination
\begin{equation}\label{eq:16}
\Theta(\x;\t) =\sum_{\alpha =0}^nt_{\alpha }\Theta ^{\alpha }(\x),\qquad t_0=1
\end{equation}%
of functions $\Theta^{\alpha }$ which satisfy the same conjugate net system \eqref{eq:13}. The critical points $\bbeta$ of $\Theta$ are then implicitly but uniquely defined by
\begin{equation}\label{eq:17}
\Theta_{\beta_i} = \sum_{\alpha =0}^{n}t_{\alpha}\Theta^\alpha_{\beta _{i}} =0,\qquad i=1,\ldots,N,
\end{equation}%
wherever the Hessian determinant $\det{(\Theta _{\beta _{i}\beta _{k}})}_{i,k}$ does not vanish. The latter is equivalent to \mbox{$\Theta_{\beta_i\beta_i}\neq 0$} since $\Theta_{\beta_i\beta_k}=0$ for $i\neq k$. Hence, in the following, we confine ourselves to the regular sector for which these conditions hold. Now, differentiation of (\ref{eq:17}) with respect to $t_\gamma$ yields
\begin{equation}\label{E5}
\Theta^\gamma_{\beta_i} + \sum_{\alpha =0}^{n}\sum_{k=1}^N t_{\alpha}\Theta^\alpha_{\beta _{i}\beta_k}\frac{\partial\beta_k}{\partial t_\gamma} =0,
\end{equation}
while differentiation of (\ref{eq:16}) with respect to $x_i$ and $x_k$ and evaluation at the critical points $\bbeta$ produce
\begin{equation}
\Theta_{\beta_i\beta_k} = \sum_{\alpha=0}^n t_\alpha\Theta^\alpha_{\beta_i\beta_k}
\end{equation}
so that changing the order of summation in (\ref{E5}) leads to
\begin{equation}\label{eq:18}
\frac{\partial \beta _{i}}{\partial t_{\gamma }}=-\frac{\Theta^\gamma_{\beta _{i}}}{\Theta _{\beta _{i}\beta _{i}}},\qquad i=1,\ldots,N,\qquad \gamma =1,\ldots,n,
\end{equation}%
where we have used the fact that $\Theta_{\beta_i\beta_k}=0$ for $i\neq k$.

The system of first-order equations (\ref{eq:18}) determines the evolution of the critical points $\bbeta$ as a function of the ``times'' $t_\gamma$. Elimination of $\Theta_{\beta_i\beta_i}$ gives rise to the system of hydrodynamic type
\begin{equation}\label{eq:19}
\frac{\partial \beta _{i}}{\partial t_{\gamma }}=\lambda _{i}^{\gamma
}(\bbeta )\frac{\partial \beta _{i}}{\partial t_{1}},\qquad i=1,\ldots,N,\qquad\gamma
=2,\ldots,n
\end{equation}%
with characteristic speeds defined by
\begin{equation}\label{eq:20}
\lambda _{i}^{\gamma }(\x)=\left.\frac{\partial \Theta^{\gamma}}{\partial x_i}\right/\frac{\partial \Theta^{1}}{\partial x_i}
\end{equation}%
and Riemann invariants $\beta _{i}$. The corresponding generalised Tsarev equations are given by (\ref{eq:17}) divided by $\Theta^1_{\beta_i}$. Differentiation of (\ref{eq:20}) shows that 
\begin{equation}\label{eq:21}
\frac{\partial \lambda _{i}^{\gamma}}{\partial x_{k}} = B_{ik}(\lambda
_{k}^{\gamma }-\lambda _{i}^{\gamma }),\qquad B_{ik} = A_{ki}\left.\frac{\partial \Theta^1}{\partial x_k}\right/\frac{\partial \Theta^{1}}{\partial x_i},\qquad i\neq k
\end{equation}
by virtue of the conjugate net equations (\ref{eq:13}). The latter also imply that
\begin{equation}
B_{ik}=\frac{\partial}{\partial x_k}\ln B_i,\qquad B_i = \frac{1}{A_i}\frac{\partial\Theta^1}{\partial x_i}
\end{equation}%
so that the definition of a semi-Hamiltonian hydrodynamic-type system
\begin{equation}\label{eq:22}
\frac{\partial B_{ik}}{\partial x_l} = \frac{\partial B_{il}}{\partial x_k},\qquad i\neq k\neq l\neq i
\end{equation}
is indeed met. Thus, any parametric family of solutions $\Theta(\x;\t)$ of the linear system \eqref{eq:13} of the form \eqref{eq:16} defines a commuting set of semi-Hamiltonian hydrodynamic-type systems \eqref{eq:19} governing the critical points $\bbeta$ of $\Theta$. Moreover, any Combescure transform 
\begin{equation}
\tilde{\Theta}(\x;\t) = \sum_{\alpha=0}^nt_\alpha\tilde{\Theta}^\alpha(\x)
\end{equation}
of $\Theta(\x;\t)$ with the Combescure transforms $\tilde{\Theta}^\alpha$ of $\Theta^\alpha$ being defined by the compatible systems
\begin{equation}
  \frac{\partial \tilde{\Theta}^\alpha}{\partial x_i} = \sigma_i\frac{\partial \Theta^\alpha}{\partial x_i},\qquad \frac{\partial \sigma_i}{\partial x_k} = A_{ik}(\sigma_k - \sigma_i),\qquad i\neq k
\end{equation}
gives rise to the same criticality conditions (\ref{eq:17}) and, hence, the same critical points $\bbeta$. Accordingly, any (equivalence) class of Combescure-related functions $\Theta$ encodes a single set of commuting hydrodynamic-type systems (\ref{eq:19}), (\ref{eq:20}).

An immediate consequence of the hydrodynamic-type system \eqref{eq:19} is that
\begin{equation}\label{eq:23}
\frac{\partial \Theta^{\alpha }(\bbeta)}{\partial t_{\gamma }}=\frac{\partial
\Theta^{\gamma}(\bbeta)}{\partial t_{\alpha }},\qquad \alpha,\gamma\geq1.
\end{equation}%
In particular, for $\gamma=1$, the conservation laws (\ref{E4b}) are retrieved and (\ref{eq:20}) coincides with (\ref{eq:8}) (modulo $\beta_i\rightarrow x_i,\,\alpha\rightarrow\gamma$). In general, (\ref{eq:23}) implies that there exists a ``potential'' function $F(\t)$ such that
\begin{equation}\label{eq:24}
\Theta^{\alpha}(\bbeta)=\frac{\partial F}{\partial t_{\alpha }}.
\end{equation}%
In fact, it is easy to verify that
\begin{equation}
F(\t) = \Theta(\bbeta(\t);\t)
\end{equation}
up to an additive constant. 

Large classes of solutions of the Darboux systems (\ref{eq:13}), (\ref{darboux}) have been constructed by, for instance, the $\partial$-bar-dressing method \cite{ZakMan85,Kon93} and  the algebro-geometric approach \cite{Kri97}. In the following, we treat the conjugate net equations (\ref{eq:13}) in a different manner.

\section{Lauricella functions and hydrodynamic-type systems}

Conjugate net equations of Euler-Poisson-Darboux-type represented by the choice
\begin{equation}\label{E6}
A_{i}\sim\prod_{k\neq
i}(x_{i}-x_{k})^{-\varepsilon _{k}}
\end{equation}
are of particular importance. The system of linear PDEs of the type \eqref{eq:3} was studied a long time ago not only in the context of classical differential geometry (see, e.g., \cite{Darboux}) but also in connection with multi-dimensional generalisations of Euler's hypergeometric function \cite{Lau}. The EPD system of equations \eqref{eq:3} has a number of remarkable properties. In
particular, its general solution may be written in the form
\begin{equation}\label{eq:26}
f(x_{1},\ldots,x_{N})=\sum_{k=1}^{N}\int_{{\mathfrak c}_{k}}f_{k}(z)\prod_{i=1}^{N}(z-x_{i})^{-\varepsilon _{i}}dz,
\end{equation}%
where $f_{k}(z)$ and ${\mathfrak c}_{k}$ are
arbitrary functions and contours of integration on the complex $z$-plane respectively (see, e.g.,
\cite{Lau, Loo, Sti}). A particular choice of functions $f_{k}(z)$ and contours ${\mathfrak c}_{k}$ gives rise to the Lauricella  function $F_{D}$ which appears in various studies. In what follows, we will refer to the functions \eqref{eq:26} as Lauricella-type functions.

A large variety of hydrodynamic-type systems \eqref{eq:19} may now be constructed by considering functions $\Theta$ which are linear combinations of Lauricella-type functions
\eqref{eq:26}. In the following, we briefly state the main results without going into details. Thus, for instance, if all functions $f_k$ vanish except for one which is of the form $ f_{k_0}=\frac{1}{2\pi i}\sum_{\alpha =1}^nt_{\alpha }z^{\alpha +\varepsilon
_{1}+\cdots+\varepsilon _{N}-1}$ and the corresponding contour ${\mathfrak c}_{k_0}$ constitutes a (counterclockwise oriented) circle of large radius then $\Theta$ may be taken to be 
\begin{equation}
\Theta(\x;\t) = \Theta^0(\x) + \mbox{\rm Res}_{z=\infty
}\left(\sum_{\alpha =1}^nt_{\alpha }z^{\alpha -1}\prod _{i=1}^{N}\left(1-\frac{x_{i}}{z}\right)^{-\varepsilon _{i}}\right),
\end{equation}
where $\Theta^0$ is an arbitrary solution of the EPD system (\ref{eq:3}). Therefore, $\Theta^\alpha$ and, by virtue of (\ref{eq:20}), the characteristic speeds $\lambda_i^\alpha$ of the corresponding hydrodynamic type system (\ref{eq:19}) are polynomials in $\x$ of degree $\alpha$ and $\alpha-1$ respectively. In the case $\varepsilon _{i}=\frac{1}{2}$, $i=1,\ldots,N$, the system \eqref{eq:19} represents the dispersionless limit of the multi-component KdV equation or its higher-order counterparts \cite{KMM}. In particular, for $N=2$,
the dispersionless NLS equation or one-layer Benney system with $\lambda_{1}=\frac{1}{2}(3x_{1}+x_{2})$, $\lambda_{2}=\frac{1}{2}(x_{1}+3x_{2})$ 
is obtained \cite{Zak80,KMM}. On the other hand, if $\varepsilon _{1}=\varepsilon _{2}=-\frac{1}{2}$ then $f(x_{1},x_{2})=-\frac{1}{2}t_{1}(x_{1}+x_{2})-\frac{1}{8}t_{2}(x_{1}-x_{2})^{2}+\cdots$ \cite{KMM11} and the characteristic speeds are $\lambda _{1}=\frac{1}{2}(x_{1}-x_{2})$, $\lambda _{2}=\frac{1}{2}(x_{2}-x_{1})$. In terms of the variable $\phi =\exp[\frac{1}{4}(\beta_{1}-\beta _{2})^{2}]$, this system is equivalent to 
\begin{equation}
\frac{\partial^2\phi}{\partial t_2^2}=\frac{\partial^2 \exp \phi}{\partial t_1^2}
\end{equation}
which is nothing but the dispersionless Toda equation \cite{Kod90} or the 1+1-dimensional Boyer-Finley equation \cite{BoyFin82}. More generally, for arbitrary $N\geq 2$ and all $\varepsilon _{i}=-\frac{1}{2}$, the $N$-component extensions of the dispersionless Toda equation are generated. In particular, for even $N$, the corresponding hydrodynamic-type systems govern deformations of the branch points for the special Seiberg-Witten spectral curves \cite{KMM13}.

The mixed cases with $\varepsilon _{i}=\pm \frac{1}{2}$ are likewise of interest. Thus, for $N=2$ and  $\varepsilon _{1}=\frac{1}{2}$, $\varepsilon _{2}=-\frac{1}{2}$, one obtains the hydrodynamic-type system
\begin{equation}
\frac{\partial \beta _{1}}{\partial t_{2}}=\frac{1}{2}(3\beta _{1}-\beta
_{2})\frac{\partial \beta _{1}}{\partial t_{1}},\qquad \frac{\partial \beta
_{2}}{\partial t_{2}}=\frac{1}{2}(\beta _{1}+\beta _{2})\frac{\partial \beta
_{2}}{\partial t_{1}}
\end{equation}
which, on applying the substitution $\beta _{2}\rightarrow -\beta _{2}$, may be regarded as a hybrid of the dispersionless NLS and Toda equations. The case $N=4$ and $\varepsilon _{1}=\varepsilon _{2}=\frac{1}{2},$ $\varepsilon_{3}=\varepsilon _{4}=-\frac{1}{2}$ leads to
\begin{equation}
\begin{array}{l}
f(x_{1},x_{2,}x_{3},x_{4})=\displaystyle t_{1}\frac{1}{2}(x_{1}+x_{2}-x_{3}-x_{4})\\[3mm]
\qquad\qquad\mbox{}+\displaystyle t_{2}\frac{1}{8}[3x_{1}^{2}+2x_{1}x_{2}+3x_{2}^{2}-2(x_{1}+x_{2})(x_{3}+x_{4})-(x_{3}-x_{4})^{2}]+\cdots
\end{array}
\end{equation}
 so that the corresponding hydrodynamic-type system with four Riemann invariants has the characteristic speeds
\begin{equation}
\begin{array}{rlrl}
\lambda _{1} =&\displaystyle\frac{1}{2}(3x_{1}+x_{2}-x_{3}-x_{4}), \qquad &
\lambda _{2} =&\displaystyle\frac{1}{2}(x_{1}+3x_{2}-x_{3}-x_{4})\\[4mm]
\lambda _{3} =&\displaystyle\frac{1}{2}(x_{1}+x_{2}+x_{3}-x_{4}), \qquad &
\lambda _{4} =&\displaystyle\frac{1}{2}(x_{1}+x_{2}+x_{4}-x_{3}).
\end{array}
\end{equation}
The latter give rise to a hydrodynamic-type system which contains the dispersionless NLS equation ($\beta_3=\beta_4=0$) and dispersionless Toda system ($\beta_1=\beta_2=0$) as particular reductions.

Characteristic speeds of an entirely different nature are obtained if one makes the choice
\begin{equation}\label{eq:27}
\Theta(\x;\t) = \Theta^0(\x) + 
f(\x;\t)=\Theta^0(\x) + t_{1}\frac{1}{2}\sum_{i=1}^{N}x_{i}+t_{2}%
\int_{x_{k}}^{x_{k+1}}\prod_{i=1}^{N}(z-x_{i})^{-\frac{1}{2}}dz
\end{equation}%
for some $k<N$. Indeed, one is led to the system
\begin{equation}\label{eq:28}
\frac{\partial \beta _{i}}{\partial t_{2}}=\lambda _{i}(\bbeta )\frac{%
\partial \beta _{i}}{\partial t_{1}},
\end{equation}%
wherein the characteristic speeds $\lambda _{i}$ are derivatives of hyperelliptic integrals, that is, 
 \begin{equation}
\lambda _{i}(\x)=2\frac{\partial}{\partial x_i}\left(\int_{x_{k}}^{x_{k+1}}\prod_{k=1}^{N}(z-x_{k})^{-\frac{1}{2}}dz\right).
\end{equation}
For $N=3,4$, such systems have been considered in \cite{Pav}. More general hydrodynamic-type systems may be constructed by choosing
\begin{equation}\label{eq:29}
\Theta = \Theta^0 + t_{1}\sum_{i=1}^{N}\varepsilon
_{i}x_{i}+t_{2}\int_{x_{k}}^{x_{k+1}}\prod_{i=1}^{N}(z-x_{i})^{-%
\varepsilon _{i}}dz.
\end{equation}

Another example of hydrodynamic-type systems with underlying Lauricella-type functions corresponds to
the choice
\begin{equation}\label{eq:30}
\Theta =\Theta^0 + t_{1}\sum_{i=1}^{N}\varepsilon _{i}x_{i}+t_{2}F_{D}(a,\varepsilon
_{1},\ldots,\varepsilon _{N},b;x_{1},\ldots,x_{N}),
\end{equation}%
where $F_{D}(a,\varepsilon _{1},\ldots,\varepsilon _{N},b;x_{1},\ldots,x_{N})$
is defined by
\begin{equation}\label{eq:Lau}
F_D:=\frac{\Gamma (b)}{\Gamma (a)\Gamma (b-a)}\int_{1}^{\infty }z^{\varepsilon
_{1}+\cdots+\varepsilon
_{N}-b-2}(z-1)^{b-a-1}\prod_{i=1}^{N}(z-x_{i})^{-\varepsilon _{i}}dz.
\end{equation}
The latter constitutes Lauricella's generalisation of the classical hypergeometric function
to the case of $N$ variables \cite{Lau}. Here, $a$ and $b$ are arbitrary
parameters and $\Gamma$ is the standard Gamma function.. The hydrodynamic-type system \eqref{eq:19}
is therefore of the form
\begin{equation}\label{eq:31}
\frac{\partial \beta _{i}}{\partial t_{2}}=\frac{1}{\varepsilon _{i}}\frac{%
\partial F_{D}(\bbeta )}{\partial \beta _{i}}\frac{\partial \beta _{i}}{%
\partial t_{1}}.
\end{equation}%
Different choices of the constants $a,\varepsilon _{1},\ldots,\varepsilon _{N},b$ give rise to various integrable hydrodynamic-type systems of interest.

\section{Whitham equations}

The relative simplicity of Lauricella-type functions make them very attractive as functions which may be employed in the classical theory of transformations of conjugate nets (see, e.g., \cite{Eisenh}). Thus, given particular  solutions of the EPD equations, one may construct large classes of more complicated systems of conjugate net equations. Specifically, gauge transformations are perhaps the simplest way of generating new systems of conjugate net equations from any given one. Indeed, if $f$ and $f^1$ are solutions of a given set of conjugate net equations (\ref{eq:2}) with coefficients $A_{ik}^0$ then one my directly verify that the gauge transformation
\begin{equation}
  f\rightarrow\Theta = \frac{f}{f^1}
\end{equation}
leads to a another system of conjugate net equations (\ref{eq:2}) with coefficients and potentials
\begin{equation}\label{E7}
A_{ik} = A_{ik}^0 - \frac{\partial}{\partial x_k}\ln f^1,\qquad A_i = \frac{A_i^0}{f^1}.
\end{equation}
In particular, if we choose the potentials
\begin{equation}
A_i^0 = \prod_{k\neq i}(x_i - x_k)^{-\varepsilon_{k}}
\end{equation}
corresponding to the EPD equations (\ref{eq:3}) then we are led to the new potentials \cite{KS}
\begin{equation}
A_i = \frac{1}{f^1}\prod_{k\neq i}(x_i - x_k)^{-\varepsilon_{k}}.
\end{equation}

In order to apply the theory developed in Section 3, we now consider the linear combination
\begin{equation}
f = \tilde{f} + t_1+ t_2 f^0,\qquad f^0 = \sum_{i=1}^{N}\varepsilon _{i}x_{i},
\end{equation}
where $\tilde{f}$ is an arbitrary solution of the EPD equations and $1$ and $f^0$ are constant and linear solutions respectively. Hence, if we superimpose the separable solutions of the EPD equations according to
\begin{equation}
f^1 = \oint_{\mathfrak c}
\prod_{i=1}^{N}(z -x_{i})^{-\varepsilon _{i}}dz
\end{equation}
for some contour $\mathfrak c$ on the complex $z$-plane then the solution $\Theta$ of the conjugate net equations with coefficients (\ref{E7}) is of the required form
\begin{equation}\label{E8}
\Theta = \Theta^0 + t_1\Theta^1 + t_2\Theta^2,\qquad \Theta^0 = \frac{\tilde{f}}{f^1},\qquad \Theta^1 = \frac{1}{f^1},\qquad \Theta^2 = \frac{f^0}{f^1}.
\end{equation}
Accordingly, the characteristic speeds $\lambda_i=\lambda_i^2$ of the associated semi-Hamiltonian hydrodynamic-type system (\ref{eq:19})$_{n=2}$ are given by
\begin{equation}\label{eq:32}
\lambda _{i} =\left.\frac{\partial \Theta^{2}}{\partial x_i}\right/\frac{\partial \Theta^{1}}{\partial x_i} = f^0 -f^1\left.\frac{\partial f^0}{\partial x_i}\right/\frac{\partial f^1}{\partial x_i}.
\end{equation}
For $N=3,4$, $\epsilon_i=\frac{1}{2}$ and an appropriate choice of the contour $\mathfrak c$, this hydrodynamic-type system coincides with the classical one-phase Whitham equations for the averaged KdV and NLS equations respectively (see, e.g., \cite{Wh,Pav2}). In the form \eqref{eq:32}, the
characteristic speeds for the one-phase Whitham equations have been given in \cite{GKE, Kud}.

In the classical differential-geometric context of conjugate nets, (\ref{eq:32}) regarded as a linear map $f^0\mapsto \lambda_i$ is known as a Levy transformation with respect to the independent variable $x_i$, generated by $f^1$ \cite{Eisenh}. This observation indicates that Levy transformations should play an important role in the theory of Whitham equations. However, it turns out that the identification of the underlying Levy transformations is not as straightforward in the case of the multi-phase Whitham equations. Thus, we  consider an $N+1$-dimensional extension of the EPD system
(\ref{eq:3}), namely
\begin{equation}\label{eq:3435}
\begin{array}{rl}
\displaystyle
\frac{\partial^2 f}{\partial x_{i}\partial x_{k}} &\displaystyle=\frac{1}{x_{i}-x_{k}}\left(\varepsilon
_{k}\frac{\partial f}{\partial x_{i}}-\varepsilon _{i}\frac{\partial f}{\partial x_{k}}\right),\\[6mm]
\displaystyle
\frac{\partial^2f}{\partial x_{i}\partial y} &\displaystyle=\frac{1}{x_{i}-y}\left(\varepsilon _{k}\frac{\partial f}{\partial x_{i}}-(1-g)\frac{\partial f}{\partial y}\right),
\end{array}
\qquad i\neq k\in\{1,\ldots,N\},
\end{equation}
 where $g$ is a positive integer and $y$ constitutes an auxiliary variable. It admits solutions of the form
\begin{equation}\label{eq:36}
f^{k}=\oint_{{\mathfrak c} _{k}}\frac{(z-y)^{g-1}}{\prod_{i=1}^{N}{(z-x_{i})}^{\varepsilon _{i}}}dz
\end{equation}%
for some contours ${\mathfrak c}_k$, $k=1,\ldots, N$. The $(g-1)$-fold Levy transformation of any solution $\varphi $ of \eqref{eq:3435} with respect to the auxiliary variable $y$ is given by (cf.\ \cite{MatSal91})
\begin{equation}\label{eq:37}
\varphi _{g-1}=\mathcal{L}_{g-1}[\varphi] =\frac{\left\vert 
\begin{array}{cccc}
\varphi & \varphi _{y} & \cdots & \varphi _{(g-1)y} \\[1mm] 
\hat{\fb} & \hat{\fb}_{y} & \cdots & \hat{\fb}_{(g-1)y}%
\end{array}%
\right\vert }{\left\vert 
\begin{array}{ccc}
\hat{\fb}_{y} & \cdots & \hat{\fb}_{(g-1)y}%
\end{array}%
\right\vert },\qquad \hat{\fb} = \left(\begin{array}{c}f^2\\ \vdots\\ f^g\end{array}\right).
\end{equation}%
It constitutes a solution of the conjugate net equations \eqref{eq:2} with appropriate coefficients $A_{ik}$ which are independent of the choice of $\varphi $. For $g=1$, the above is to be interpreted as $\varphi_0 = \varphi$.

The Levy transforms $f^0_{g-1} = \mathcal{L}_{g-1}[f^0]$ and $f^1_{g-1} = \mathcal{L}_{g-1}[f^1]$ of the two particular solutions
\begin{equation}\label{eq:38}
f^{0}=\sum_{i=1}^{N}\varepsilon _{i}x_{i}+(1-g)y, \qquad f^{1}
\end{equation}%
of the extended EPD system (\ref{eq:3435}) 
therefore read
\begin{equation}\label{eq:39}
f_{g-1}^{0}=\frac{\sum_{i=1}^{N}\varepsilon _{i}x_{i}\left\vert 
\begin{array}{ccc}
\hat{\fb}_{y} & \cdots & \hat{\fb}_{(g-1)y}%
\end{array}%
\right\vert +(g-1)\left\vert 
\begin{array}{cccc}
\hat{\fb}-y\hat{\fb}_{y} & \hat{\fb}_{yy} & \cdots & \hat{\fb}_{(g-1)y}%
\end{array}%
\right\vert }{\left\vert 
\begin{array}{ccc}
\hat{\fb}_{y} & \cdots& \hat{\fb}_{(g-1)y}%
\end{array}%
\right\vert }
\end{equation}%
and
\begin{equation}\label{eq:40}
f _{g-1}^{1}=\frac{\left\vert 
\begin{array}{ccc}
\fb & \cdots & \fb_{(g-1)y}%
\end{array}%
\right\vert }{\left\vert 
\begin{array}{ccc}
\hat{\fb}_{y} & \cdots & \hat{\fb}_{(g-1)y}%
\end{array}%
\right\vert },\qquad \fb = \left(\begin{array}{c} f^1\\ \vdots\\ f^g\end{array}\right).
\end{equation}%
Since the functions $f^{k}$ are polynomials in $y$, one concludes that
\begin{equation}\label{eq:41}
f_{g-1}^{0}=\frac{\sum_{i=1}^{N}\varepsilon _{i}x_{i}\left\vert 
\begin{array}{ccc}
\hat{\mbox{\boldmath $I$}}_{g-2} & \cdots& \hat{\mbox{\boldmath $I$}}_{0}%
\end{array}%
\right\vert -\left\vert 
\begin{array}{cccc}
\hat{\mbox{\boldmath $I$}}_{g-1} & \hat{\mbox{\boldmath $I$}}_{g-3} & \cdots & \hat{%
\mbox{\boldmath $I$}}_{0}%
\end{array}%
\right\vert }{\left\vert 
\begin{array}{ccc}
\hat{\mbox{\boldmath $I$}}_{g-2} & \cdots & \hat{\mbox{\boldmath $I$}}_{0}%
\end{array}%
\right\vert }
\end{equation}%
and
\begin{equation}\label{eq:42}
f_{g-1}^{1}=\frac{\left\vert 
\begin{array}{ccc}
{\mbox{\boldmath $I$}}_{g-1} & \cdots & {\mbox{\boldmath $I$}}_{0}%
\end{array}%
\right\vert }{\left\vert 
\begin{array}{ccc}
\hat{\mbox{\boldmath $I$}}_{g-2} & \cdots & \hat{\mbox{\boldmath $I$}}_{0}%
\end{array}%
\right\vert },
\end{equation}%
where the components $I_l^k$ of $\mbox{\boldmath $I$}_{l}$ and $\hat{\mbox{\boldmath $I$}}_{l}$ are given by
\begin{equation}\label{eq:43}
I_{l}^k=\oint_{{\mathfrak c}_{k}}\frac{z
^{l}}{\prod_{i=1}^{N}{(z -x_{i})}^{\varepsilon _{i}}}dz.
\end{equation}
By construction, the functions $f_{g-1}^{0}$ and $f_{g-1}^{1}$ are solutions of the same
conjugate net equations \eqref{eq:2}. It is important to note that these solutions do not depend on
the auxiliary variable $y$. This is due to the fact that the particular solutions (\ref{eq:38}) are polynomials in $y$ of degree less than $g$. 

Now, consistent with the interpretation of the $g-1$-fold Levy transformation  (\ref{eq:37}) for $g=1$ as the identity transformation, we observe that $f^0_0 = f^0$ and $f^1_0=f^1$. Hence, the canonical  generalisation of the function $\Theta$ given by (\ref{E8}) reads
\begin{equation}\label{eq:44}
\Theta = \Theta^0 + t_1\Theta^1 + t_2\Theta^2,\qquad \Theta^0 = \frac{\tilde{f}_{g-1}}{f^1_{g-1}},\qquad \Theta^1 = \frac{1}{f^1_{g-1}},\qquad \Theta^2 = \frac{f^0_{g-1}}{f^1_{g-1}},
\end{equation}
where $\tilde{f}_{g-1}$ is an arbitrary solution of the conjugate net equations generated by the iterated Levy transformation $\mathcal{L}_{g-1}$. As pointed out in the preceding, since $f_{g-1}^0, f_{g-1}^1,\tilde{f}_{g-1}$ and $1$ are all solutions of the same conjugate net equations, the ratios $\Theta^0,\Theta^1$ and $\Theta^2$ obey another system of conjugate net equations. Hence, once again, we may apply the theory developed in Section 3 and state that the functions 
\begin{equation}\label{eq:45}
\lambda _{i} =\left.\frac{\partial \Theta^{2}}{\partial x_i}\right/\frac{\partial \Theta^{1}}{\partial x_i} = f^0_{g-1} -f^1_{g-1}\left.\frac{\partial f^0_{g-1}}{\partial x_i}\right/\frac{\partial f^1_{g-1}}{\partial x_i}
\end{equation}
constitute characteristic speeds of an associated semi-Hamiltonian hydrodynamic-type system (\ref{eq:1}). The nature of the latter depends on the choice of the parameters $\varepsilon_i$. The characteristic speeds $\lambda_i$ may be interpreted as the Levy transforms of $f^0_{g-1}$ with respect to the independent variables $x_i$. In a different context, the action of Levy transformations on hydrodynamic-type systems has been investigated in detail in \cite{Fer}.

Remarkably, for the particular choice $N=2g+1$, $\varepsilon _{i}=\frac{1}{2}$ and the contours $%
{\mathfrak c}_k$ being canonical cycles for the associated Riemann surfaces of
genus g, the system \eqref{eq:1} with characteristic speeds \eqref{eq:45} is nothing but the
classical $g$-phase Whitham system for the KdV equation. Indeed, if we set
\begin{equation}\label{eq:46}
H_{1}=\frac{\left\vert 
\begin{array}{ccc}
\hat{\mbox{\boldmath $I$}}_{g-2} & \cdots & \hat{\mbox{\boldmath $I$}}_{0}%
\end{array}%
\right\vert }{\left\vert 
\begin{array}{ccc}
{\mbox{\boldmath $I$}}_{g-1} & \cdots & {\mbox{\boldmath $I$}}_{0}%
\end{array}%
\right\vert },\qquad H_{2}=-\frac{\left\vert 
\begin{array}{cccc}
\hat{\mbox{\boldmath $I$}}_{g-1} & \hat{\mbox{\boldmath $I$}}_{g-3} & \cdots & \hat{%
\mbox{\boldmath $I$}}_{0}%
\end{array}%
\right\vert }{\left\vert 
\begin{array}{ccc}
{\mbox{\boldmath $I$}}_{g-1} & \cdots & {\mbox{\boldmath $I$}}_{0}%
\end{array}%
\right\vert }
\end{equation}%
and $\Gamma _{0}=1$, $\Gamma_{1}=\frac{1}{2}\sum_{i=1}^{2g+1}x_{i}$ then
\begin{equation}\label{eq:47}
\lambda _{i} =\left.\frac{\partial(H_{1}\Gamma _{1}+H_{2}\Gamma _{0})}{\partial x_i}\right/ \frac{\partial H_1}{\partial x_i},\qquad i=1,\ldots,2g+1
\end{equation}%
which coincides with the expressions for the characteristic speeds given in~\mbox{\cite{Tian, El}}. Other choices of the seed solution $f^{0}$ give rise to other members of Whitham hierarchies, all of which describe deformations of hyperelliptic curves \mbox{$p^{2}=\prod_{i=1}^{2g+1}(z-\beta _{i})$}. An extension of these results to the case $N=2g+2$ is straightforward. The corresponding equations \eqref{eq:1}, \eqref{eq:45} coincide with the $g$-phase Whitham equations for the NLS equation \cite{FL}. Hydrodynamic-type systems \eqref{eq:1}, \eqref{eq:45} with $\varepsilon _{i}=1/K$ govern the deformation of superelliptic  $(K,N)$ curves $p^{K}=\prod_{i=1}^{N}(z-\beta _{i})$. Finally, we note that our approach to Whitham type equations appears to be different from that used in \cite{OS} to construct hydrodynamic-type systems via generalised hypergeometric functions.

\section*{Acknowledgements}

The first author (YK) was partially supported by NSF grants DMS-1108813 and DMS-1410267.  The second author (BK) acknowledges support by the PRIN 2010/2011 grant 2010JJ4KBA\_003.

\end{document}